\documentclass[12pt]{article}
\usepackage{graphicx}
\usepackage{xspace}
\usepackage{bm}
\usepackage{color}

\usepackage[pdftex,
    plainpages=false,
    bookmarks         = true,
    bookmarksnumbered = true,
    pdfpagemode       = None,
    pdfstartview      = FitH,
    pdfpagelayout     = SinglePage,
    colorlinks        = true,
    linkcolor         = blue,
    pagecolor         = black,
    urlcolor          = blue,
    pdfborder         = {0 0 0}
    ]{hyperref}%

\definecolor{blue}{rgb}{0,0,0}
\definecolor{green}{rgb}{0,0,0}
                  
\definecolor{ourEm}{rgb}{1,0,0}

\newcommand\pubnumber{TUM-HEP-794/11}
\newcommand\pubdate{\today}

\newcommand{\beq}{\begin{equation}}
\newcommand{\eeq}{\end{equation}}

\newcommand{\unit}[1]{\ensuremath{\rm\,#1}}

\newcommand{\TeV}{\unit{TeV}}

\newcommand{\ps}{\unit{ps}}

\newcommand{\Dzero}{D{\O}}

\def\address{1. Physics Department,
University of Warwick, United Kingdom \\
2. Physik Department, Technische Universit{\"a}t M{\"u}nchen,
D-85748 Garching, Germany \\
 3. CPPM Aix-Marseille Universit\'e IN2P3 CNRS, France}

\def\Title#1{\begin{center} {\Large #1 } \end{center}}
\def\Author#1{\begin{center}{ \sc #1} \end{center}}
\def\Address#1{\begin{center}{ \it #1} \end{center}}

\newcommand\pubblock{\rightline{\begin{tabular}{l} \pubnumber\\
         \pubdate  \end{tabular}}}
\newenvironment{Abstract}{\begin{quotation}  }{\end{quotation}}
\newenvironment{Presented}{\begin{quotation} \begin{center} 
             PRESENTED AT\end{center}\bigskip 
      \begin{center}\begin{large}}{\end{large}\end{center} \end{quotation}}
\def\Acknowledgements{\bigskip  \bigskip \begin{center} \begin{large}
             \bf ACKNOWLEDGEMENTS \end{large}\end{center}}
\newcommand{\niceMKbabar}{\mbox{\sl B\hspace{-0.4em}
{\small\sl%
A}\hspace{-0.37em} \sl B\hspace{-0.4em} {\small\sl%
A\hspace{-0.02em}R}}}

\newcommand{\Jpsi}{\ensuremath{J\!/\!\psi}\xspace}

\begin{document}
\begin{titlepage}
\pubblock

\vfill
\Title{Summary of WG4 \\ ``Lifetime, mixing and weak mixing
phase in charm and beauty, including direct determination of
$V_{tx}$''}
\vfill
\Author{Michal Kreps$^1$, Alexander Lenz$^2$, Olivier
Leroy$^3$}
\Address{\address}
\vfill
\begin{Abstract}
We present the summary of the Working Group on lifetimes,
mixing and weak mixing phases in charm and beauty mesons at
the CKM 2010 workshop. In the past year or so good progress was
achieved on both experimental and theoretical {
sides}. While
this yields improvement { in} our understanding of neutral meson
mixing, further work is necessary to achieve the highest
possible precision in order to investigate current
hints for deviations between experiment and standard model predictions. 
With the recent LHC startup we see bright
prospects for the near term future for huge improvements. 
\end{Abstract}
\vfill
\begin{Presented}
6th International Workshop on the CKM Unitarity Triangle \\
University of Warwick, United Kingdom, September 6-10, 2010 
\end{Presented}
\vfill
\end{titlepage}
\def\thefootnote{\fnsymbol{footnote}}
\setcounter{footnote}{0}
%


\section{Introduction}

Mixing of neutral mesons provided an important tool for the
development of the standard { model and continues to be important
to test it.} 
It also provides crucial
information in searches for  physics beyond the standard model
and constraining models of new physics. In this paper we
concentrate on the mixing of neutral mesons containing
$b$ or $c$ quarks. In the bottom and charm sectors,
improvements on both experiment and theory
{sides} since
the previous CKM workshop brought advances in the quest for 
new physics. In addition with the start of the LHC operation we
are {entering a new era} in which tests with unprecedented
precision will  become reality.

In this paper we summarise 13 contributions to the working
group on ``Lifetime, mixing and weak mixing
phase in charm and beauty'' together with lively discussions
triggered by those contributions. They covered all aspects
of mixing of $b$ and $c$ mesons as well as prospects for
the future improvement of our knowledge of mixing.

\section{$\mathbf{B}$ mixing}


\subsection{Mixing formalism}

\noindent
We start with the description of mixing of $B_q$ mesons ($q=d,s$), 
which is governed by the Schr\"odinger-like equation
\begin{equation}
i \frac{d}{dt}
\left(
\begin{array}{c}
| B_q(t) \rangle \\ | \bar{B}_q (t) \rangle
\end{array}
\right)
=
\left( \hat{M}^q - \frac{i}{2} \hat{\Gamma}^q \right)
\left(
\begin{array}{c}
| B_q(t) \rangle \\ | \bar{B}_q (t) \rangle
\end{array}
\right),
\end{equation}
where $\hat{M}^q$ and $\hat{\Gamma}^q$ are mass and decay
rate $2\times 2$ hermitian matrices.
The box diagrams for $B_q$
mixing give rise to off-diagonal elements
$M^q_{12}$ and $\Gamma_{12}^q$ in the mass matrix
$ \hat{M}^q$ and the decay rate matrix $ \hat{\Gamma}^q$.
Diagonalisation of   $ \hat{M}^q$ and $ \hat{\Gamma}^q$
yields mass eigenstates
\begin{eqnarray}
B_{q,H}&=&p \; B_q - q \; \bar{B}_q,  \\
B_{q,L}&=&p \; B_q + q \; \bar{B}_q,  
\end{eqnarray}
with $p$ and $q$ being complex numbers satisfying $|p|^{2} +
|q|^{2}  =  1$. The off-diagonal elements of the mass and
decay matrices can be related to three measurable
observables: the mass difference 
\begin{equation}
\Delta M_q := M_H^q - M_L^q  = 
        2 {{ |M_{12}^q|}} \left( 1 +
        { \frac{1}{8} \frac{|\Gamma_{12}^q|^2}{|M_{12}^q|^2} \sin^2 \phi_q +
...}\right),
\end{equation}
the decay width difference
\begin{equation}
\Delta \Gamma_q := \Gamma_L^q - \Gamma_H^q = 
        2 { |\Gamma_{12}^q| \cos  \phi_q }
        \left( 1 -
        { \frac{1}{8} \frac{|\Gamma_{12}^q|^2}{|M_{12}^q|^2} \sin^2 \phi_q 
        + ...}\right)
\end{equation}
and the {CP asymmetry in flavour specific
decays}
\begin{eqnarray}
      {a_{fs}^q} & = &   \mbox{Im} \frac{\Gamma_{12}^q}{M_{12}^q} { + {\cal O} \left( \frac{\Gamma_{12}^q}{M_{12}^q} \right)^2 } 
                 = \frac{\Delta \Gamma_q}{\Delta M_q} \tan \phi_q { + {\cal O} \left( \frac{\Gamma_{12}^q}{M_{12}^q} \right)^2 }, 
\end{eqnarray}
where $\phi_q = \mbox{arg}( -M_{12}^q/\Gamma_{12}^q)$.

\subsection{Standard model predictions for the mixing quantities}

\noindent
In the standard model the off-diagonal elements $M_{12}^q$
are given by
\begin{eqnarray}
{M_{12}^q} & = & \frac{G_F^2}{12 \pi^2} 
          { (V_{tq}^* V_{tb})^2} M_W^2 S_0(x_t)
          { B_{B_q} f_{B_q}^2}  M_{B_q}
\hat{\eta }_B \,.
\end{eqnarray}
$\Gamma_{12}^d$ is negligibly small compared to the current
experimental precision. On the contrary $\Gamma_{12}^s$ is large enough to be
important. The standard model predicts for $\Delta \Gamma_s$ 
\cite{ln06}
\begin{eqnarray}
\Delta \Gamma_s & = & 
{ \left( \frac{f_{B_s}}{240 \, \mbox{MeV}}
\right)^2}
\left[ { 0.105 { B}} + { 0.024  {
\tilde B_S'}} - 
0.027  { B_R} \right],
\\
\frac{\Delta \Gamma_s}{\Delta M_s}  & = & \, \, \, \,
\, \, \,  \, \, \, \, \, \, \, \,  \, \, \, \, \, \, 
\, \, \, \, \, \, \, \, \, \, \, \, \, 
\left[ { 46.2}  \, \,  \, \, \, \,  \, \, \, \,
+ { 10.6 { \frac{ \tilde B_S'}{B}} } 
- 11.9 { \frac{B_R}{B}}
\right]\times 10^{-4}.
\end{eqnarray}
In these expressions, $V_{tq}$ and $V_{tb}$ are CKM matrix
elements, $M_W$ and $M_{B_q}$ are masses of $W$ boson and
$B_q$ meson, $S_0(x_t)$ and $\hat{\eta }_B$ include
the perturbative part and finally $B_{B_q}$, $f_{B_q}$, $B$,
$B_S'$ and $B_R$ contain the non-perturbative part of the
amplitude. The CKM
matrix elements together with the non-perturbative contribution
are currently the two dominant factors limiting the precision of
the standard model predictions.
The non-perturbative matrix elements are evaluated using 
non-perturbative methods like lattice
QCD, which made significant progress over the past few years
{\cite{Shigemitsu:2011vi}}.
As an example the precision on $f_{B_d}$ and $f_{B_s}$ is now in
the region of 4--8\% with further prospects for improvements.
The reassuring fact of those calculations is that results of
unquenched calculations from
several collaborations are now available and in good
agreement. In addition, different collaborations use
different formulations for heavy and light quarks, which
adds to the confidence in the values of non-perturbative
matrix elements and their uncertainties. More details are
discussed by N.~Garron in these proceedings \cite{garron}, see also
\cite{lattice} and references therein.

One point worth {noting} is that for $B_d$ and $B_s$ mesons
$\Gamma_{12}^s/M_{12}^s$ is about $5\times 10^{-3}$ 
and therefore in the expressions presented
here one can safely neglect terms containing
$(\Gamma_{12}^q/M_{12}^q)^2$.
Plugging in the latest values for all input parameters, the standard
model predictions are \cite{nierste}
\begin{eqnarray}
        \Delta M_s                       & = & (17.3 \pm 2.6) \; \mbox{ps}^{-1},
         \\
        \frac{\Delta \Gamma_s}{\Gamma_s} & = &  { 0.137 \pm 0.027},
         \\
        a_{fs}^s & =&   \left(1.9  \pm 0.3 \right) \times 10^{-5},
         \\
        \phi_s   & = & 0.22^\circ \pm 0.06^\circ,
         \\ 
        \frac{\Delta \Gamma_d}{\Gamma_d} & = & \left( 4.2 \pm 0.8 \right) \times 10^{-3},
                \\
        a_{fs}^d & = & - \left(4.1 \pm 0.6 \right) \times 10^{-4},
         \\
        \phi_d  & = & {-4.3^\circ} \pm {1.4^\circ},
         \\
        A_{fs}^b & = & 0.494 a_{sl}^s + 0.506 a_{sl}^d = (-2.0 \pm 0.3 ) \times 10^{-4}.
\end{eqnarray}
Here $A_{fs}^b$ is the flavour-specific asymmetry
averaged over $B^0$ and $B_s$ where {the} two weights are given by
the product of the fragmentation fraction of $b$-quarks into
given hadrons and the time integrated mixing probability \cite{Nakamura:2010zzi}.
Experimentally it can be accessed by measuring the asymmetry in
same-sign dileptons at hadron colliders to which we will turn later.

It is interesting to note that due to the progress in the lattice determination
of the decay constants, currently the dominant uncertainty in $\Gamma_{12}$ stems
from the non-perperturbative matrix elements of power-suppressed four-quark operators 
\cite{nierste}.

\subsection{Testing the HQE through lifetimes}

\noindent
The above predictions rely on the Heavy Quark Expansion (HQE),
which itself needs to be tested. One of the most accurate
tests we have these days uses lifetime ratios of the $b$ hadrons.
The lifetimes are governed by the $b\rightarrow c$
tree level transition and thus expected to be completely
dominated by the standard model. Thus if a discrepancy between
theory and experiment exists, it is a clear signal of issues
with the HQE rather than a sign of new physics. 
In the ratio of lifetimes the overall $m_b^5$-dependence as well as
several hadronic uncertainties cancel. Currently the accuracy 
of the theory predictions is strongly limited by the lack
of up-to-date values for the bag parameters of the arising 
non-perturbative four-quark matrix elements.
Using the ten year old values of Ref. \cite{baglife}
one obtains \cite{nierste}
\begin{eqnarray}
\frac{\tau (B_s)}{\tau (B_d)} -1       & \in & 
[-4 \times 10^{-3}; 0  ],
\\
\frac{\tau (B^+)}{\tau (B_d)} - 1      & = & 0.044 \pm 0.024.
\end{eqnarray}
{Here the $\tau (B_s)$ is defined as inverse of the mean decay width of the two mass eigenstates.} For the error estimates all numerical
input parameters were varied within their one sigma range 
and the individual uncertainties were finally added quadratically.
We emphasize again that the numerical value of the lifetime predictions
depends strongly on the values of the color-suppressed four-quark matrix elements, that
are hardly known, see \cite{nierste} for more details.
A typical number quoted for the $\Lambda_b$ lifetime is e.g. \cite{Tarantino:2007nf}
\begin{eqnarray}
\frac{\tau (\Lambda_b)}{\tau (B_d)}  & = & 
0.88 \pm 0.05.
\end{eqnarray}
It is worth to note that the theory prediction for $\Lambda_b$ is not
as complete as for the other two ratios. 
Currently we still lack the full NLO-QCD calculation and the input 
from lattice QCD. 

Despite that $B^0$ and $B^+$ lifetimes were known at
subpercent precision from $B$-factories, CDF recently joined
the game and provided new measurements, which match the precision of
the $B$-factories and are consistent with them \cite{malde}. The world average
lifetimes are $\tau(B^+)=1.638 \pm 0.011$\ps\ and
$\tau(B^0)=1.525 \pm 0.009$\ps\ \cite{hfag}. This translates to the ratio
$\tau(B^+)/\tau(B^0)=1.074\pm 0.010$ \cite{hfag} which is consistent
with theory. The measurements of the $B_s$ lifetimes are more
difficult due to the non-zero decay width difference. Best
measurements come from the angular analysis of
$B_s\rightarrow\Jpsi\phi$ decays where one can measure
$\tau(B_s)=2/(\Gamma_{qH}+\Gamma_{qL})$. 
The latest results are $\tau(B_s)={1.529} \pm 0.025 \pm 0.012$
ps from CDF \cite{kreps} and $\tau(B_s)=1.45 \pm 0.04 \pm 0.01$\ps\ from
\Dzero\ \cite{borissov}. Our average of those two measurements yields the
lifetime ratio $\tau(B_s)/\tau(B^0)-1=(-13.1\pm 5.9)\times
10^{-3}$, again consistent with theory expectation.
The $\Lambda_b$ lifetime is the most problematic part. On
one hand the theory prediction is incomplete. On the other
hand the experimental results are not in good agreement. The
latest two measurements from CDF which are of highest
precision yield results which are well above all other
measurements. In numbers, CDF measures 
$\tau(\Lambda_b)=1.401\pm 0.046\pm 0.035$ \ps\ using the
$\Lambda_b\rightarrow \Lambda_c\pi$ decay
and $\tau(\Lambda_b)=1.537\pm0.045\pm0.014$ \ps\ using
the $\Lambda_b\rightarrow\Jpsi\Lambda$ decay \cite{malde}. Performing a 
naive
average which neglects correlated uncertainties the lifetime
from CDF is $\tau(\Lambda_b)=1.483\pm0.037$\ps\ while the world
average excluding the CDF measurements is
$\tau(\Lambda_b)=1.230\pm 0.074$\ps. Also using the two most
precise measurements the ratio of $\Lambda_b$ to $B^0$
lifetimes is above the prediction, but given that the theory
prediction is incomplete and the discrepancies on the experimental
side we should not draw any conclusion about the validity of
HQE from that yet.

{We complete this section by discussing  the
decay width difference in the $B_s$ system.} Both CDF and \Dzero\
measure together with the mean $B_s$ lifetime also
$\Delta\Gamma_s$. Omitting details discussed elsewhere in
these proceedings, they obtain values of $\Delta\Gamma_s=0.075
\pm 0.035 \pm 0.01$\,ps$^{-1}$ (CDF) \cite{kreps} and $\Delta\Gamma_s=0.15
\pm 0.06 \pm 0.01$\,ps$^{-1}$ (\Dzero) \cite{borissov}. While both have central
values in the region of theory expectations, the precision is not
yet sufficient to firmly establish a non-zero decay width
difference {nor} to obtain a strong test of theory.
{Contrary to the $B_s$ system in $B^0$ system
the decay width difference does not get lot of attention. As
pointed out by T.~Gershon experiments should turn back to
question of the decay width difference in the $B^0$ system
\cite{Gershon:2010wx}.}

\section{Direct determination of $\mathbf{V_{td}}$, $\mathbf{V_{ts}}$ and $\mathbf{V_{tb}}$}

The elements of the CKM matrix are free parameters of the standard
model and as such they need to be extracted from experiments.
Ideally one would like to have a determination which is
independent of assumptions on the underlying physics. In
practice it is non-trivial to determine $V_{tx}$ elements
with sufficient precision without some assumptions. 

In the context of the standard model the main feature of the CKM
matrix is its unitarity, which reduces the number of free
parameters to four. Using the absolute values of the elements in
first two rows supplemented by the single phase $\gamma$ measured in
$B\rightarrow D^{(*)}K$ decays with the assumption of unitarity
it is possible to extract $V_{tx}$ elements from tree level
processes with astonishing precision. Further improvements
in precision can be achieved by {including} loop level
processes into {the} determination. A current analysis using only
tree level quantities yields \cite{utfit}
\begin{eqnarray}
V_{td} &=& (0.00896\pm\;0.0006\;\;{\&}\;\; 0.01081\!\pm\!
0.0006)\!\times\!e^{i(-22.9\pm 1.4)^\circ}, \\
V_{ts} &=& -0.03979\pm\! 0.00052\!\times\! e^{i(-1.163\pm
0.084)^\circ}, \\
V_{tb} &=& 0.99916\pm\! 1.8\!\times\!10^{-5}.
\end{eqnarray}
{The two values of $V_{td}$ stem from two independent values
of the CKM angle $\gamma$ used in the fit.}
Once we drop the requirement of unitarity, the situation is more
difficult as the well measured elements from the first two rows do not
provide strong constraints anymore. As an example fourth
generation or models with additional vector-like quarks are
often discussed. In the first case, the mixing matrix becomes a
$4\times 4$ matrix while in the second case it becomes a
$4\times 3$ matrix. In such cases all elements can get
modified, but with the high precision direct measurements of
the first two rows, most of the impact is in the $V_{tx}$ elements.
In both cases large modifications of $V_{ts}$ and $V_{td}$ are
possible. For additional
discussion we refer the reader to the discussion by Rohrwild~\cite{rohrwild}
or the original papers, e.g. Ref.~\cite{SM4}.

Experimentally we can access the $V_{tx}$ elements without
imposing an unitarity constraint in top quark physics. Both
$t\bar{t}$ pair production as well as
electroweak single top quark production can be used.  
The $V_{tx}$ elements enter those processes both at
the production as well as at the decay. 
The Tevatron experiments used both production
{mechanisms}. When
analysing the $t\bar{t}$ pair production sample, they split the data
according to the number of $b$-tagged jets. In the fit they extract 
in addition to the cross section itself also
$\mathcal{R}_b=|V_{tb}|^2/(|V_{tb}|^2+|V_{ts}|^2+|V_{td}|^2)$.
The \Dzero\ analysis yields $\mathcal{R}_b=0.97^{+0.09}_{-0.08}$
including both statistical and systematic uncertainties
\cite{Abazov:2008yn}.
Using the 95\% CL limit $\mathcal{R}_b > 0.79$ from
\cite{Abazov:2008yn} leads to the
constraint 
$$
|V_{td}|^2+|V_{ts}|^2<0.263\cdot|V_{tb}|^2 \,.
$$
The single-top cross-section is measured by both CDF
\cite{Aaltonen:2009jj} and \Dzero\ \cite{Abazov:2009ii} and
their combined result is 
$\sigma=(2.76^{+0.58}_{-0.47})\,\mu$b. Comparison of this
result with cross-section predictions 
which use $|V_{tb}|=1$ yields
$|V_{tb}|=0.88\pm0.07$. Clearly the precision of those
determinations is not very high. Moreover the  Tevatron
experiments are already close to {being} systematically limited and
therefore, despite the much larger statistics expected at LHC, it
will require non-trivial work to substantially improve those
results. For more a detailed discussion of the experimental
aspects see Ref.~\cite{wagner}.

\section{New physics in $\mathbf{B}$ meson mixing}

\subsection{Model independent analysis}

\noindent
An important thread for the current studies of $B$ meson mixing
is the quest for  new physics beyond the standard model. Being a loop
induced process, it is a well suited laboratory for such
searches. For the $B_s$ system new physics can be 
parametrised as \cite{ln06}
\begin{eqnarray}
\Gamma_{12,s} =  \Gamma_{12,s}^{\rm SM}\, ,
&&
M_{12,s}  =  M_{12,s}^{\rm SM} \cdot { \Delta_s} \, ;
\, \, \, \, \, \, \,  
{\Delta_s} = { |\Delta_s|} e^{i { \phi^\Delta_s}}. 
\end{eqnarray}
It should be noted that in this parametrisation new
physics does not affect $\Gamma_{12}$. While this is not
strictly correct, for most of the models typically discussed
the space for new physics to affect $\Gamma_{12}$ is very
limited. This limitation comes from the lifetime of 
$B_s$ mesons, which would be affected by such a contribution.
The general agreement is that a new physics contribution to
$\Gamma_{12}$ is well below the hadronic uncertainties and
is thus usually omitted.
With this parametrisation the observables become
\begin{eqnarray}
 \Delta M_s  & = & 2 | M_{12,s}^{\rm SM} | \cdot { |\Delta_s |},
\\
\Delta \Gamma_s  & = & 2 |\Gamma_{12,s}|
\cdot \cos \left({ \phi_s^{\rm SM} + {
\phi^\Delta_s} }\right),
\\
a_{fs}^s 
&= &
{  \frac{|\Gamma_{12,s}|}{|M_{12,s}^{\rm SM}|}} 
\cdot \frac{\sin \left( { \phi_s^{\rm SM} + {
\phi^\Delta_s} }\right)}
{ |\Delta_s|},
\\
{ \phi_s^{\Jpsi \phi}} & = & 
{ -2 \beta_s + \phi_s^\Delta + \delta_{\rm
Peng.}^{\rm SM}  + \delta_{\rm Peng.}^{\rm NP}}.
\end{eqnarray}
The quantity $\phi_s^{\Jpsi \phi}$ is the $CP$ violating phase
measured in $B_s\rightarrow \Jpsi\phi$ decays with
$\beta_s=\arg\left(-V_{ts}V_{tb}^*/V_{cs}V_{cb}^*\right)$.
It can receive the standard model contribution from the tree
level decay diagram with interference with and without
mixing ($-2\beta_s$), the contribution from new physics in the
box diagram ($\phi_s^\Delta$) and the contribution from
penguin decay diagrams both from the standard model and  new
physics, see \cite{ciuchini} for a discussion of the standard model penguins
in the case of $B_d$ decays.

\subsection{New physics in $\mathbf{B_d}$ mixing}

\noindent
{Let us first discuss
the $B^0_d$ system, which currently shows an about $2.5\sigma$
discrepancy between the $CP$ violation in $B^0_d\rightarrow\Jpsi K_S$ measured by the experiments and
a one determined from the unitarity triangle fits which omits it on input
\cite{lnckmf,Charles:2004jd,Lunghi:2009ke,Bevan:2010gi}.}
The final measurement from the BABAR experiment yields
$\sin(2\beta)=0.687\pm 0.028\pm 0.012$ \cite{Aubert:2009yr}.
The {most recent} measurement
of the Belle experiment gives
$\sin(2\beta)=0.650\pm0.029\pm0.018$ \cite{Chen:2006nk}. Belle still analyses
its full dataset with improvements to the software, which
offers improvements better than just scaling with
statistics. With the final dataset they expect to achieve
a statistical uncertainty on $\sin(2\beta)$ of $0.024$.
The value extracted by the UTFit collaboration when omitting
$\sin(2\beta)$ from the fit is $0.771\pm0.036$ which is about
$2.6\sigma$ away from the world average of measurements
\cite{utfit}. A similar
conclusion holds also for other unitarity triangle fits
\cite{Charles:2004jd,Lunghi:2009ke}. 
One of the questions is whether this discrepancy is an early
indication of new physics or whether the penguin contributions
which are typically neglected can account for this
discrepancy. This question was discussed in the presentation by
M.~Ciuchini~\cite{ciuchini} who presented several ways of assessing 
the effects of
the penguin pollution. From his analysis it seems unlikely
that a significant part of the discrepancy would be due to
neglected penguin pollution. With the estimates which are
possible with current data, the discrepancy can be decreased to
the level of $2.3\sigma$.

\subsection{New physics in the $\mathbf{B_s}$ system}

\noindent
The most important measurement these days
in terms of the search for new physics in $B_s$ mixing is
the measurement of the $CP$ violating phase $\phi_s^{\Jpsi \phi}$.
Both Tevatron experiments performed this measurement at the
end of 2007 and early 2008 for the first time
\cite{Aaltonen:2007he,Abazov:2008fj}. At that time
both experiments had result where the consistency between data
and the standard model was about $1.5\sigma$, which caused quite
some excitement.
\begin{figure}[tbh]
\centering
\includegraphics[width=6.0cm]{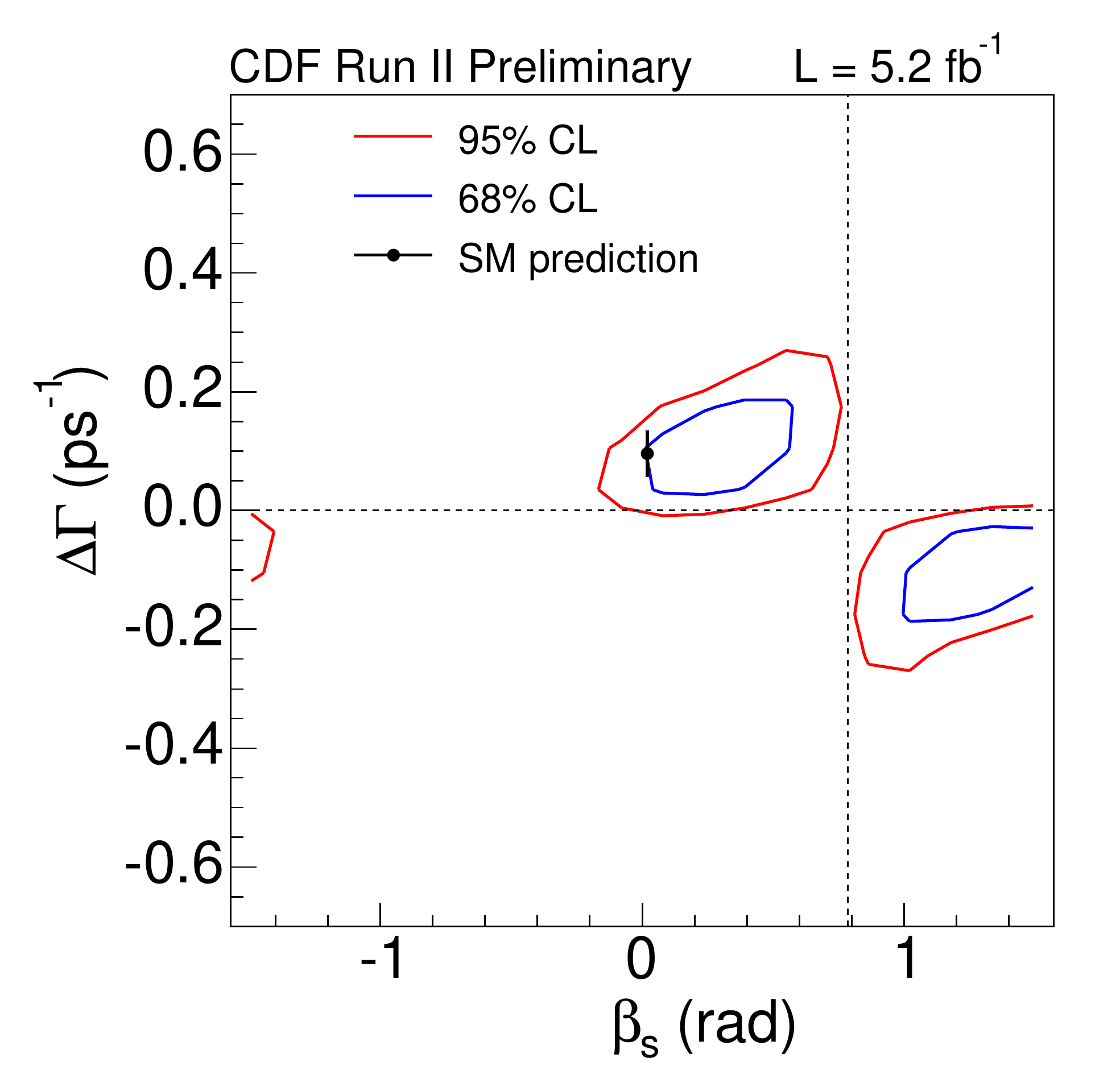}
\includegraphics[width=7.2cm]{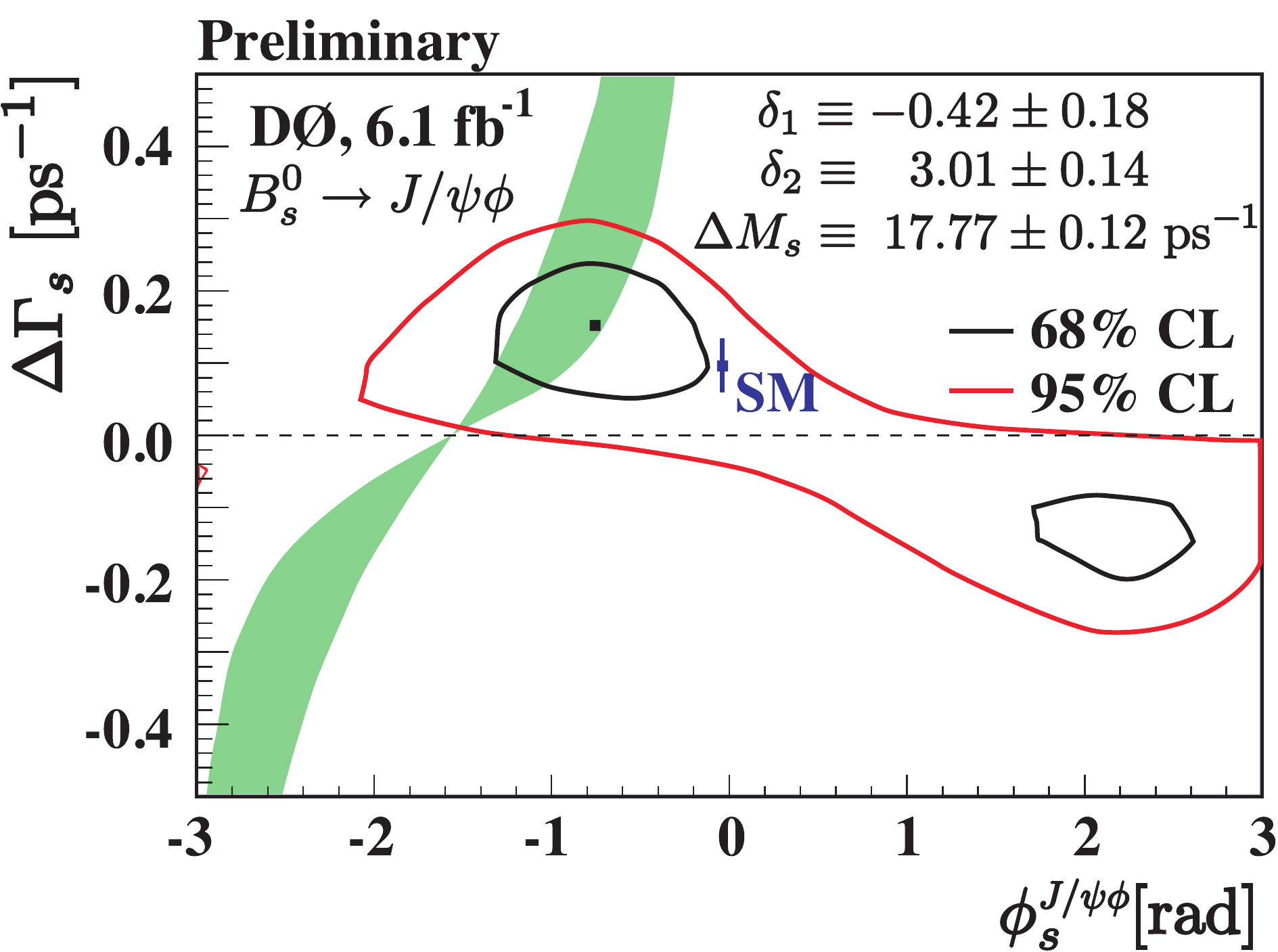}
\includegraphics[width=6.2cm]{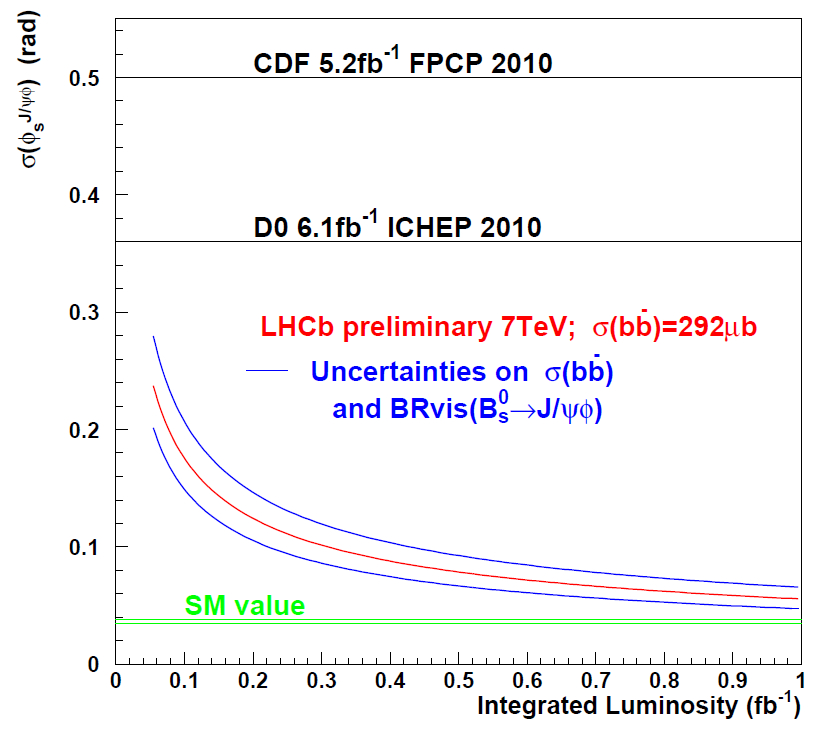}
\caption{Confidence regions in
$\Delta\Gamma_s$-($\phi_s^{\Jpsi \phi}\equiv-2\beta_s$) plane
measured in $B_s\rightarrow\Jpsi\phi$ decays at CDF (left
left) and \Dzero\ (top right). Projection of expected uncertainty on
$\phi_s^{\Jpsi \phi}$ at the LHCb experiment as a function of
integrated luminosity for running at $\sqrt{s}=7$\TeV\
(bottom).}
\label{fig:betasTevatron}
\end{figure}
In 2010 both experiments updated their analyses, CDF using
5.2 fb$^{-1}$ \cite{kreps} and \Dzero\ using 6.1 fb$^{-1}$
\cite{borissov}. The extracted
confidence regions in $\Delta\Gamma_s$-$\phi_s^{\Jpsi \phi}$
plane are shown in Fig.~\ref{fig:betasTevatron}.
Projecting down the CDF result yields $\phi^{\Jpsi\phi} \in
[-1.04,\, -0.04] ~ \cup ~ [-3.10,\, -2.16]$ rad at 68\%CL. The \Dzero\ experiment
extracts a value $\phi^{\Jpsi\phi} = -0.76^{+0.38}_{-0.36} ({\rm
stat}) \pm 0.02({\rm syst})$  rad. It should be noted that
both experiments now see better agreement between the standard
model and their data, but large new physics effects cannot
be excluded.
While the Tevatron experiments still collect data and expect
about a factor of 2 more by the end of 2011, the future of this
measurement is at the LHCb experiment. With only 600\,nb$^{-1}$
they could extract first $B_s\rightarrow\Jpsi\phi$ signal.
While the simulation does not yet fully agree with the early data, none
of the discrepancies significantly limits the capability of the
LHCb experiment to perform the analysis. The Monte Carlo projections shown in
Fig.~\ref{fig:betasTevatron}  indicate that LHCb can be competitive with
data taken in 2011~\cite{stephie}, but one should be a little cautious as
the uncertainty 
depends on the values of other physics
parameters like $\Delta\Gamma_s$.

The second measurement sensitive to new physics in
the $B_{(s)}$ mixing phase is the measurement of flavour-specific
asymmetries $a_{fs}^q$. Despite {the fact} that these asymmetries are 
expected to be small even in the case of large new physics effects, 
the \Dzero\ experiment
managed to perform this challenging measurement by measuring
the asymmetry between same sign dimuons in $b$ events.
They measure $A_{sl}^b$ which is a mixture of $a_{fs}$ for
$B^0$ and $B_s$ to be $(-95.7\pm25.1\pm14.6)\times 10^{-4}$
\cite{borissov,Abazov:2010hj},
which is about $3.2\sigma$ away from the standard model
expectation of $(-2.3^{+0.5}_{-0.6}) \times 10^{-4}$
\cite{ln06}. {The updated SM prediction of $A_{sl}^b$ announced at the CKM workshop \cite{nierste}
has no visible effect on the significance of the discrepancy.} If this stands, this is probably the strongest 
indication of new physics {in a particle
physics experiment} we have. 
While no other measurements which would be competitive to the \Dzero\
one are available up to now, LHCb presented an interesting idea
of a complementary measurement. 
The plan is to measure the semileptonic asymmetry  separately for 
 $B^0_d$ and $B_s$, using the same final state  $K K\pi l\nu$.
This way, the detector asymmetry cancels in the difference. 
It opens up
the possibility for a precise measurement which would be
complementary to the \Dzero\ measurement. Combination of the two
measurements would allow to
disentangle contributions from $B_s$ and $B^0$ with high
precision. With LHCb taking data and the Tevatron experiments
having a large sample already available we should see progress
very soon.

A combined fit of observables both in the $B_d^0$ as well as in the
$B_s$-system tends to favor new physics acting in both systems, see e.g.
\cite{nierste,lnckmf}. 

\section{Charm sector}

The mixing of charm mesons provides information
complementary to $B$ mesons or kaon mixing. The reason for this
stems from the fact that while in $B$ mesons and kaons up
type quarks run in the loops, in the charm system down type
quarks are contributing to the loops. The phenomenology is
in principle the same as for $B$ mixing, but here
$\Gamma_{12}/M_{12}$ is of order one which requires to
consider the exact formulas for $\Delta M (M_{12}, \Gamma_{12})$ 
and for  $\Delta \Gamma (M_{12}, \Gamma_{12})$, compared to the
approximate ones given in Section~2.
While several calculations for the mixing parameters $x=\Delta
M/\Gamma$ and $y=\Delta\Gamma/2\Gamma$ exist \cite{charm-mixing}, 
a satisfactory approach does not exist yet \cite{petrov}. 
One of the main difficulties
comes from the fact that the short-range contribution to $D$
mixing is strongly suppressed by the GIM mechanism combined
with a strong Cabbibo suppression of the $b$ quark in the loop,
while long range contributions might be sizeable. While precise
predictions are difficult, $x$ and $y$ up to order of 1\%
are not excluded in the standard model.

Charm mixing poses also difficulties on the experimental
side despite huge samples collected by the experiments. In
fact while $D^0$ mixing is established with more than
10$\sigma$ significance, no single measurement above
5$\sigma$ exists. Difficulties come from very slow mixing,
when the majority of the produced $D^0$ mesons decay before having
a significant chance to oscillate, and a rather small decay width
difference, which require extremely good control over
systematic effects in order to establish a significant
difference. More details of the current results and
experimental status are discussed by 
Malde~\cite{malde} and Meadows~\cite{meadows}.

Since quite some time  large $CP$ violation in the charm
sector {has been} 
considered as a smoking gun for new physics. Independent of
the type of  $CP$ violation, thanks to the large Cabbibo
suppression of the $b$-quark contribution in charm processes
it is almost impossible to generate large $CP$ violation in
the standard model. Different authors put their upper bounds
to slightly different values, but a general consensus is that
$CP$ violation in the charm sector can be at most few times
$10^{-3}$ in the standard model. So if $CP$ violation of a few
percent is measured by the experiments it would provide a clear
signal of new physics. Several experimental searches for $CP$
violation exist, but up to now no significant effect is
seen, {neither} for mixing induced nor for direct $CP$ violation. For
direct $CP$ violation, the most precise experimental
information exists for decays of $D^0$ to two charged pions
or kaons. The world average asymmetries are
{$A_{CP}(\pi^+\pi^-)=0.002\pm0.04$ and
$A_{CP}(K^+K^-)=-0.0017\pm0.0031$ \cite{Nakamura:2010zzi}}. A couple of weeks after
the conference, a new measurement by CDF was released with
$A_{CP}(\pi^+\pi^-)=0.0022\pm0.0026$ \cite{cdfD0ACP}, which will further
improve the world average. {It is worth noting that 
in $\pi^+\pi^-$ and $K^+K^-$ final states experiments
now enter the region where the standard model explanation cannot be
excluded even if $CP$ violation is observed.}

It should be noted that despite the difficulties in
the calculations, $D$ mixing already provides strong bounds on
some new physics models. The sensitivity reaches up to scales of
$10^2$--$10^3$\TeV\ in a case of natural couplings, or
{indicates} suppressed couplings in case of new physics at 1\TeV\ scale,
see e.g~\cite{petrov,NPincharm}. 

While current experiments analysed almost all their data,
with LHCb and future $B$-factories there are good prospects
for further improvements. With just a couple of months of
physics running the LHCb experiment could clearly
demonstrate their capability of collecting charm samples
\cite{gersabeck}.
One of the first public results measuring charm hadron
cross-sections shows good agreement with theory and
demonstrates the capabilities of the experiment \cite{gersabeck}. 
Signals of many $D$ meson decay modes were already
established. As an example with only 124\,nb$^{-1}$ a signal of
about 680 $D^{*+}\rightarrow D^0\pi^+$ with $D^0\rightarrow
K_s\pi^+\pi^-$ is established. The main unknown for LHCb is
the question how well the systematic uncertainties can be controlled
and how much triggers will be suppressed with increasing
luminosity. We refer the reader to Ref.~\cite{gersabeck} for more
information.

It should be also noted that measurements of  charm mixing in
different decay modes do not provide directly $x$ and $y$ but
their linear combinations determined by the strong phase
involved in process. The strong phase itself can be
extracted in experiments at $D\bar{D}$ threshold like CLEO-c
or BES III. Without those experiments future improvements in
$D$ mixing will soon become limited without real benefit of
a huge statistics from LHCb or future $B$-factories
\cite{meadows}.

\section{Conclusions}

In summary good progress in $B$, $D$ mixing, lifetimes and
the determination of the $V_{tx}$ elements of the CKM matrix
{has been} made.
Many new results are available and especially the ones in the
$B_s$ sector cause excitement in the community. While
the current generation of  experiments comes to the end 
a lot of useful data is still to be analysed. 
LHC has started its operation with all LHC experiments
including LHCb clearly demonstrating their capabilities and
readiness for physics. This together with positive
decisions on the future generation $e^+e^-$ $B$-factories
provides bright prospects for the future on the experimental side. On
the theory side a large community works on the topics discussed
here with gradual progress on all fronts including hard work
towards more precise predictions within the standard model.
Altogether the authors see great prospects for a lot of progress
over the next two years and expect exciting sessions at the next
CKM workshop. 

\Acknowledgements

The authors would like to thank the organisers of the CKM
workshop for the very good organisation and the fruitful 
physics discussions held during our stay in the
University of Warwick. 




\end{document}